\documentclass[a4paper,11pt]{article}
\pdfoutput=1 % if your are submitting a pdflatex (i.e. if you have
\usepackage{jcappub} % for details on the use of the package, please
\usepackage[T1]{fontenc} % if needed

\usepackage{subfigure}
\usepackage{enumitem}
\usepackage{pifont}
\setlist[itemize]{itemsep=0.5ex,parsep=0pt,label=\checkmark}

\usepackage{color, colortbl}
\usepackage[dvipsnames]{xcolor}
\definecolor{Gray}{gray}{0.9}
\definecolor{Gray2}{gray}{0.95}
\definecolor{LightCyan}{rgb}{0.88,1,1}

\usepackage{subfigure}

\usepackage{cancel}

\usepackage[normalem]{ulem}

\newcommand{\blueeq}[1]{{\textcolor{blue}{#1}}}

\title{\boldmath Toward a test of Gaussianity of a gravitational wave background}

\author[a,1]{Reginald Christian Bernardo\note{Corresponding author}}
\author[a,b]{, Stephen Appleby}
\author[c,d]{, and Kin-Wang Ng}
\affiliation[a]{Asia Pacific Center for Theoretical Physics, \\ Pohang 37673, Korea}
\affiliation[b]{Department of Physics, POSTECH, \\ Pohang 37673, Korea}
\affiliation[c]{Institute of Physics, Academia Sinica, \\ Taipei 11529, Taiwan}
\affiliation[d]{Institute of Astronomy and Astrophysics, Academia Sinica, \\ Taipei 11529, Taiwan}
\emailAdd{reginald.bernardo@apctp.org}
\emailAdd{stephen.appleby@apctp.org}
\emailAdd{nkw@phys.sinica.edu.tw}
\abstract{
The degree of Gaussianity of a field offers insights into its cosmological nature, and its statistical properties serve as indicators of its Gaussianity. In this work, we examine the signatures of Gaussianity in a gravitational wave background (GWB) by analyzing the cumulants of the one- and two-point functions of the relevant observable, using pulsar timing array (PTA) simulations as a proof-of-principle. This appeals to the ongoing debate about the source of the spatially-correlated common-spectrum process observed in PTAs, which is likely associated with a nanohertz stochastic GWB. We investigate the distribution of the sample statistics of the one-point function in the presence of a Gaussian GWB. Our results indicate that, within PTAs, one-point statistics are impractical for constraining the Gaussianity of the nanohertz GWB due to dominant pulsar noises. However, our analysis of two-point statistics shows promise, suggesting that it may be possible to constrain the Gaussianity of the nanohertz GWB using PTA data. We also emphasize that the Gaussian signatures identified in the one- and two-point functions in this work are expected to be applicable to any gravitational wave background.
}

\begin{document}
\maketitle
\flushbottom

\section{Introduction}
\label{sec:introduction}

The first direct detection of a gravitational wave (GW) has ushered in a new era in astronomy \cite{LIGOScientific:2016aoc}. Hundreds of GW events have now been observed from Solar mass compact objects \cite{KAGRA:2021vkt}, each one carrying important information about the strong gravity regime \cite{LIGOScientific:2021sio}. The next breakthrough in GW astronomy is the detection of a gravitational wave background (GWB) \cite{Caprini:2018mtu, Christensen:2018iqi, Romano:2019yrj, Moore:2021ibq, NANOGrav:2020spf, Pol:2022sjn, Staelens:2023xjn}, coming from a superposition of GWs from sources that are too weak to be resolved individually. Efforts are ongoing in order to resolve the GWB in the sub-kiloherz band using ground-based GW detectors \cite{LIGOScientific:2009qal, Shoemaker:2019bqt, Lehoucq:2023zlt} and in the millihertz band in future space-based GW observations \cite{LISACosmologyWorkingGroup:2022kbp, Cheng:2022vct, Liang:2021bde, Muratore:2023gxh}. Recently, pulsar timing array (PTA) collaborations have reported a compelling evidence in favor of the presence of a nanohertz stochastic GWB \cite{NANOGrav:2023gor, Reardon:2023gzh, Antoniadis:2023lym, EPTA:2023fyk, Xu:2023wog}, consistent with a common-spectrum process across pulsars with the signature Hellings and Downs (HD) correlation \cite{Hellings:1983fr}. This signal is traditionally interpreted as coming from a population of supermassive black hole binaries (SMBHB) \cite{Sazhin:1978myk, Detweiler:1979wn, Phinney:2001di, Wyithe:2002ep, Sesana:2004sp, Sesana:2008mz, Burke-Spolaor:2018bvk, Sato-Polito:2023gym, Sato-Polito:2023spo, Bi:2023tib, Sato-Polito:2024lew, Raidal:2024odr, Bernardo:2024bdc}. However, the PTAs' observation also allows for an interesting set of sources of the nanohertz GWB, rooted cosmologically \cite{Chen:2019xse, Ellis:2020ena, NANOGrav:2021flc, Buchmuller:2021mbb, Xue:2021gyq, Sharma:2021rot, Figueroa:2023zhu, Ellis:2023dgf, Saeedzadeh:2023biq, Huang:2023chx, Ye:2023tpz, Zhu:2023lbf, Jiang:2023gfe, Jiang:2024dxj, Winkler:2024olr}. A challenge to the field now is to identify the signatures in the observation that can distinguish the different sources of a GWB, ultimately settling the issue of whether an observed GWB signal is astrophysical or cosmological.

Different sources of a GWB produce the same HD spatial correlation, only with a different spectrum. A logical step to determine the source of the the signal is thus by calculating the expected spectrum for each source and computing the respective evidences relative to the data. This approach has been widely adopted by the PTA community, but has not ruled out any significant possibilities, given the present data \cite{EPTA:2023xxk, NANOGrav:2023hvm, Ellis:2023oxs}. The field is optimistic that forthcoming iterations of the data {\cite{Siemens:2013zla, Moore:2014eua, Vigeland:2016nmm, Hazboun:2019vhv}}, with more and longer timed pulsars, may eventually enable PTAs to contribute to the source debate. This poses the question whether there are further untapped aspects of the observation that may be able to help out in distinguishing the source of an observed GWB signal, not only in PTA but in general. This brings us to the theme of this work which is testing the Gaussianity of a GWB.

In this work, we bare down signatures of the GWB signal in the one- and two-point functions' cumulants, referring to PTA simulations \cite{Babak:2024yhu} as a proof-of-concept. In particular, a Gaussian GWB requires that all the higher-point information and the cumulants of the observable reduce to products of the power spectrum, and the spatial correlation. The relation this has on the source is that the level of Gaussianity that is expected to be carried by the signal is different for astrophysical/SMBHBs and cosmological sources, with the latter expected to display a higher degree of Gaussianity. Testing the Gaussianity of the signal thus boils down to measuring the cumulants of a relevant observable, and checking if there is additional information in the cumulants that is not already present in the power spectrum.

This direction we embark on is reminiscent of tests of isotropy and statistics in the cosmic microwave background (CMB) \cite{Planck:2019evm}. In CMB analysis, the one-point information is constrained to assess if the data shows any significant departures from Gaussianity. If there is, then a primordial theory may be realized to try to explain why the signal exhibits such non-Gaussianity (or some late time effect such as lensing). However, we shall soon find out that the PTA/CMB analogy ends here; because in the CMB, the signal is larger compared to instrumental noise, whereas in PTAs, the GWB signal is subdominant to the intrinsic pulsar red noises {\cite{NANOGrav:2023hde, EPTA:2023sfo, Zic:2023gta}}. This compels us to venture into the signatures of a Gaussian GWB in two-point function statistics, utilizing both the power spectrum and the spatial correlation in the signal. Referring to our PTA simulations with only one hundred pulsars, this has shown some promise of not only resolving the HD signal in the mean statistic, but most importantly also the signatures of Gaussianity in the cumulants of the two-point function.

This work proceeds as follows. In Section \ref{sec:gwb_stats}, we first revisit the meaning of a Gaussian stochastic field and a Gaussian GWB (Sections \ref{subsec:gaussian_statistics}-\ref{subsec:the_gaussian_gwb}). We follow this up by elaborating on aspects of the one-point function that manifests said Gaussianity of a GWB signal in PTA simulations (Sections \ref{subsec:mock_data}-\ref{subsec:one_point_statistics}). However, as we have teased out already, in PTAs, the one-point function statistics is not going to be able to resolve the Gaussianity information in the cumulants because of the noise in pulsars are dominant compared to the signal. We then move to the two-point function which turns out to be able to potentially display the signatures of a Gaussian GWB (Section \ref{sec:two_point_statistics}). Appendix \ref{sec:clt_and_large_scale_modes} dwells into an intricacy between the central limit theorem and large scale modes. Our codes and notebooks that can be used to reproduce the main results of this work can be found in \href{https://github.com/reggiebernardo/notebooks/tree/main/supp_ntbks_arxiv.2407.17987}{GitHub}.

\section{Gravitational wave background statistics}
\label{sec:gwb_stats}

We briefly revisit the notion of a Gaussian field (Section \ref{subsec:gaussian_statistics}), a Gaussian GWB (Section \ref{subsec:the_gaussian_gwb}) and study its implications for the one-point function statistics in PTA (Sections \ref{subsec:mock_data}-\ref{subsec:one_point_statistics}).

\subsection{Gaussian statistics}
\label{subsec:gaussian_statistics}

A Gaussian stochastic field, $\Psi(x)$, is completely specified by a two-point function, $\langle \Psi(x) \Psi(x') \rangle${, a one-point (that can always be shifted to zero), and higher-point functions that admit the relations described below} \cite{Weinberg:2008zzc}. The odd moments of the field vanishes,
\begin{equation}
\label{eq:odd_moments_gaussian}
    \langle \Psi(x) \rangle = \langle \Psi(x) \Psi(x') \Psi(x'') \rangle = \cdots = 0 \,,
\end{equation}
and the even moments factorize into a product-sum of two-point functions, $\langle \Psi(x) \Psi(x') \rangle$; the four-point function becomes {\cite{Isserlis_Theorem}}
\begin{equation}
\label{eq:4p_gaussian}
\begin{split}
    & \langle \Psi_1 \Psi_2 \Psi_3 \Psi_4 \rangle 
    = \langle \Psi_1 \Psi_2 \rangle \langle \Psi_3 \Psi_4 \rangle + \langle \Psi_1 \Psi_3 \rangle \langle \Psi_2 \Psi_4 \rangle + \langle \Psi_1 \Psi_4 \rangle \langle \Psi_2 \Psi_3 \rangle \,,
\end{split}
\end{equation}
where we write down $\Psi_i = \Psi(x_i)$ for brevity. The six-point, eight-point, and higher-point functions follow suit. For this work, we shall find the following two-point identities particularly useful \cite{Srednicki:1993ix, Gangui:1994wh},
\begin{equation}
\label{eq:6p_gaussian}
    \langle \Psi_1^3 \Psi_2^3 \rangle = 9 \langle \Psi_1^2 \rangle \langle \Psi_2^2 \rangle \langle \Psi_1 \Psi_2 \rangle + 6 \langle \Psi_1 \Psi_2 \rangle^3 \,,
\end{equation}
and
\begin{equation}
\label{eq:8p_gaussian}
    \langle \Psi_1^4 \Psi_2^4 \rangle = 9 \langle \Psi_1^2 \rangle^2 \langle \Psi_2^2 \rangle^2 + 72 \langle \Psi_1^2 \rangle \langle \Psi_2^2 \rangle \langle \Psi_1 \Psi_2 \rangle^2 + 24 \langle \Psi_1 \Psi_2 \rangle^4 \,.
\end{equation}

In a Gaussian field, all information in the higher-point functions and cumulants can be unpacked into the two-point function. Thus, the Gaussianity of a field can be probed by measuring its moments. In practice, an observation $O$ can be taken as $f\left[ \Psi(x) \right]$ where $f$ is some functional of the field $\Psi(x)$. If $\Psi(x)$ is a Gaussian field, and $f$ is a linear functional, then the observation $O$ can be expected to inherit the Gaussianity of the field; that is, the statistics of $O$ will be completely specified by a two-point function $\langle O(x) O(x') \rangle$, and its higher moments will factorize as (\ref{eq:odd_moments_gaussian}-\ref{eq:8p_gaussian}) with $O(x)$ in the place of the field $\Psi(x)$.

This is the idea behind the statistics and isotropy tests of the CMB \cite{Planck:2019evm}, where $\Psi(x) \equiv \delta (x)$ are scalar density fluctuations, and $O(x) \equiv \Delta T(x)$ is the temperature. In PTA science, $\Psi(x) \equiv h(x)$ are gravitational waves, and $O(x) \equiv r(x)$ are pulsar timing residuals. There are practical limits to this analogy between the CMB and the GWB in GW detectors such as PTAs. For the CMB, the presence of power at all angular scales, and subdominant instrumental noise, enable one-point statistics to be measured to sufficient accuracy to test the Gaussianity of the signal. For PTAs, neither is true; loud pulsar noises are generally present and the GWB power is concentrated in large scale modes. We consider the implications of this in the following sections.

\subsection{The Gaussian GWB}
\label{subsec:the_gaussian_gwb}

We consider {a stationary,} isotropic and Gaussian GWB \cite{Hellings:1983fr, Phinney:2001di},
\begin{equation}
\label{eq:gaussian_sgwb}
    \langle h_\blueeq{A}\left(f, \hat{k}\right) h_\blueeq{A'}^*\left(f', \hat{k}'\right) \rangle = {\cal P}(f) \blueeq{ \delta_{AA'} } \delta\left(f - f'\right) \delta\left(\hat{k} - \hat{k}'\right) \,,
\end{equation}
where $h_\blueeq{A}\left(f, \hat{k}\right)$ are GW amplitudes{, $A=+,\times$ are GW polarization indices}, $f$ is a GW frequency with a unit wavevector $\hat{k}$, and ${\cal P}(f)$ is the power spectrum. The assumption of Gaussianity implies that all the higher point information and the cumulants of the field in the GWB can be described by ${\cal P}(f)$. Observational departures from Gaussianity can be probed through the cumulants of the relevant observable \cite{Planck:2019evm}.

In weak fields relevant to PTA, we can expect that the statistics of the GWB are going to be inherited by the pulsar timing residuals. This allows an analytical route to express the cumulants of the pulsar timing residuals' one-point function, and the corresponding cosmic variances, solely in terms of the angular power spectrum \cite{Srednicki:1993ix, Gangui:1994wh}. For our discussion, we would sometimes refer to the cosmic variance \cite{Roebber:2015iva, Roebber:2016jzl, Allen:2022dzg, Bernardo:2022xzl, Bernardo:2023bqx, Wu:2024xkp} as the ensemble variance, and the distribution of the values in different simulations/universes to be an ensemble distribution.

Throughout this work, we quantify the statistics using the sample mean (${\cal E}_1$), variance (${\cal V}_2$), skewness (${\cal S}_3$), and kurtosis (${\cal K}_4$) of a set of points/pixels on the sphere, $s\equiv \{s_1, s_2, \cdots, s_N\}$, as ${\cal E}_1 [ s ] = \sum_{i} s_i/N$, ${\cal V}_2 [ s ] = {\cal E}_1 \left[ \left( s - {\cal E}_1 [ s ] \right)^2 \right]$, ${\cal S}_3 [ s ] = {\cal E}_1 \left[ \left( s - {\cal E}_1 [ s ] \right)^3 \right]$, and ${\cal K}_4 [ s ] = {\cal E}_1 \left[ \left( s - {\cal E}_1 [ s ] \right)^4 \right]$\footnote{We are loosely interchanging the terms for the skewness and the kurtosis with the third and fourth centralized moments. Our results hold for either statistical description.}. {The quantity $s$ represents the pulsar timing residuals (sum $i$ over $n_{\rm psrs}$ pulsars, Section \ref{subsec:one_point_statistics}) or the residual-pair product (sum $i$ over $n_{\rm psrs} (n_{\rm psrs}-1)/2$ pulsar pairs, Section \ref{sec:two_point_statistics}).} The validity of the one-point statistics following \cite{Srednicki:1993ix, Gangui:1994wh} can be tested by numerically simulating the ensemble distribution of the sample statistics. Our analysis has shown that the analogous results of \cite{Srednicki:1993ix, Gangui:1994wh} for PTA holds; however, since the ensemble distribution of the one-point sample statistics turns out to be generally non-Gaussian distributed due to the spatial correlation, the first two moments, the mean and the cosmic variance, are no longer able to give a faithful picture of the true distribution (Appendix \ref{sec:clt_and_large_scale_modes}). We shall show that this is the case for both the GWB signal and red noise in PTA simulations, but with differing natures of induced non-Gaussianities.

\subsection{Mock PTA data}
\label{subsec:mock_data}

{We generate our PTA simulations following \cite{Babak:2024yhu}. This gives timing data for each pulsar in the form
\begin{equation}
\label{eq:residual_fourier_series}
r(t, \hat{e}_a)= \sum_k \left( \alpha_{ak} \sin(\omega_k t) + \beta_{ak} \cos(\omega_k t) \right) \,,
\end{equation}
where $\omega_{k}= 2\pi k f_1 =2\pi k/T$, $T$ is the span of the observation in years, $k=1,2,3,\cdots$, and $\hat{e}$ is a unit vector pointing to the direction of the pulsar relative to Earth. As a reference, for $T=1$ yr, the fundamental frequency $f_1=1/T=31.8$ nHz; for $T=30$ yr, $f_1\sim 1$ nHz. PTAs are thus able to gain access to lower frequencies through longer decades of observation.}

{For red noise, the frequency components are drawn from a Gaussian distribution, $\alpha_{ak}, \beta_{bk} \sim {\cal N}\left(0, S_{ab}(f_k) \Delta f_k \right)$ where $S_{ab}(f)=\delta_{ab} {\cal S}^{\rm N}(f)$ is the noise power spectrum and $\Delta f_k= f_{k+1}-f_k$ \cite{vanHaasteren:2014qva}. {Note that we use subscripts $a,b$ to label pulsars and $k$ for frequency bins.} We consider a simple power law for the red noise spectrum. In this case, the noise components are uncorrelated across different pulsars. For the GWB, the data is obtained by $\alpha_{ak}, \beta_{bk} \sim {\cal N}\left(0, S_{ab}(f_k) \Delta f_k \right)$ where $S_{ab}(f)=\Gamma_{ab} {\cal S}^{\rm C}(f)$, $\Gamma_{ab}$ is the HD correlation, and ${\cal S}^{\rm C}(f)$ is the power spectrum of circular SMBHBs. In this case, the GWB signal is viewed as a common spectrum that is spatially correlated across pulsars, in accordance with the HD curve.}

\subsection{One-point statistics}
\label{subsec:one_point_statistics}

\begin{figure}[h!]
    \centering
    \includegraphics[width=0.8\textwidth]{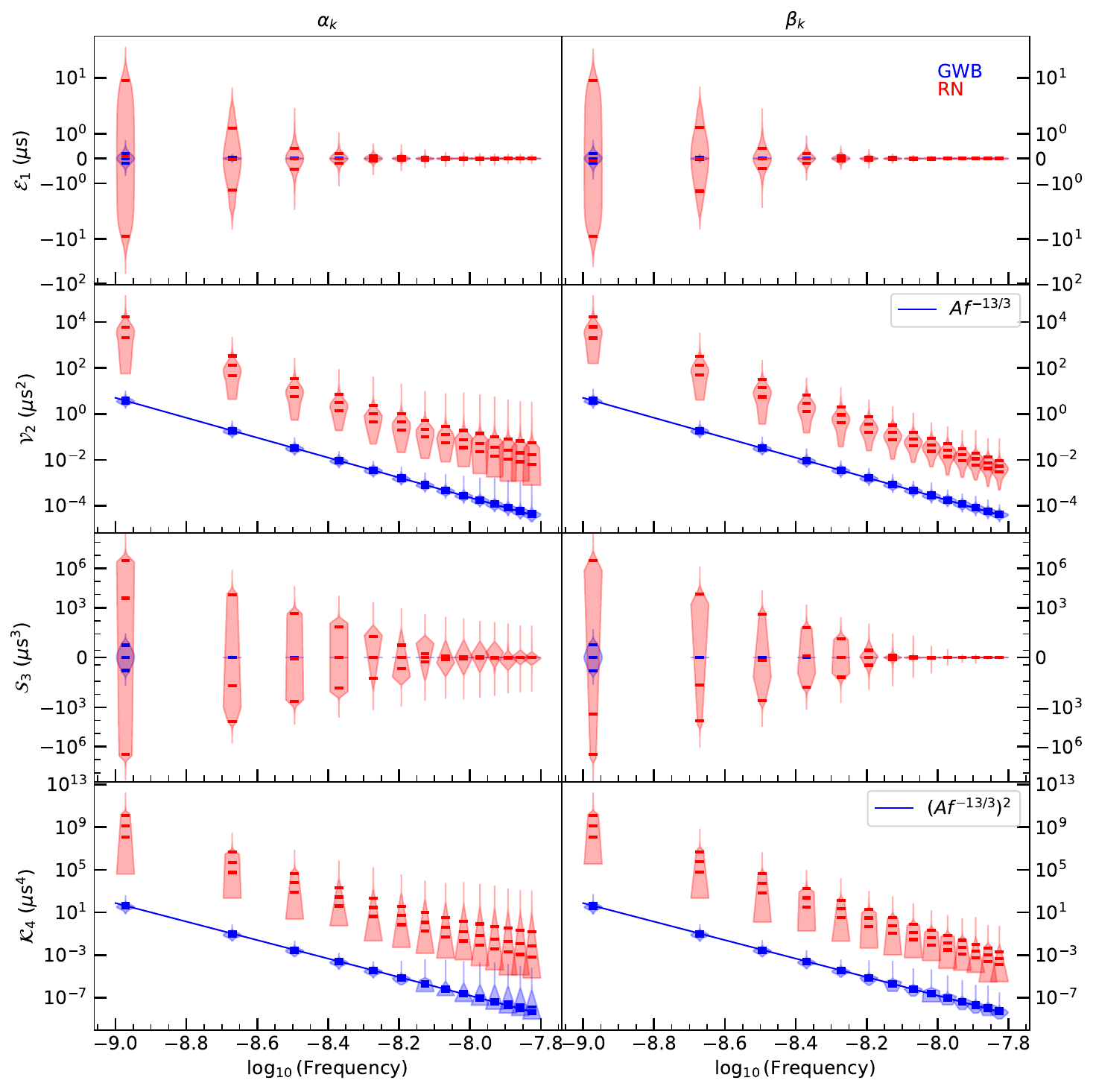}
    \caption{Ensemble distributions of the first four moments/one-point metrics of PTA timing residuals, given only GWB and red noise (RN) components; $n_{\rm psrs}=100$ pulsars and $n_{\rm sims}=5000$ simulations; red points/violins have both GWB and RN.}
    \label{fig:ensemble_stats_all_bins}
  \end{figure}

For the GWB, we consider an amplitude $A_{\rm gw}=2.4 \times 10^{-15}$ and spectral index $\gamma_{\rm gw}=13/3$ corresponding to circular SMBHBs \cite{Phinney:2001di}, along with the HD correlation. For the red noise (RN), considered as independent Gaussian random processes with a power law spectrum, we uniformly draw the amplitudes, $\log_{10}A_{\rm rn}={\cal U}(-17, -13)$, and spectral indices, $\gamma_{\rm rn}={\cal U}(2, 6)$, for each pulsar. {This gives a time series data $r(t, \hat{e}_a)$, for $a=1, 2, \cdots n_{\rm psrs}$ pulsars, and the Fourier bins $\alpha_{ak}, \beta_{ak}$ (via a standard discrete Fourier transform routine, see Eq. \eqref{eq:residual_fourier_series}). Then, the sample statistics of $\alpha_{ak}, \beta_{ak}$ are computed over the simulated PTA (one realisation); average over pulsars in the PTA. The entire process is repeated numerous times with the same signal and noise priors, each realisation turning in values of the sample statistics representative of an underlying ensemble PDF/cosmic variance, as shown in Figure \ref{fig:ensemble_stats_all_bins}.}

Figure \ref{fig:ensemble_stats_all_bins} shows the ensemble distributions (5000 samples/simulations) of the sample statistics, mean (${\cal E}_1$), variance (${\cal V}_2$), skewness (${\cal S}_3$), and kurtosis (${\cal K}_4$), of the one-point function of pulsar timing residuals in 30 yr-PTA simulations with 100 pulsars, scattered anisotropically \cite{Babak:2024yhu}.

Starting with the GWB signal, our first comment goes to the variance across frequency bins. This follows ${\cal V}_2 \sim f^{-13/3}$ as shown, reflective of the spectrum corresponding to the input GWB. In relation, the fourth centralized moment, which is proportional to the kurtosis, admits ${\cal K}_4 \sim \left( A f^{-13/3} \right)^2$. This is owed to the Gaussian nature of the GWB, such that the higher moments of the underlying field factorize into a product-sum of two-point functions. In the astrophysical setting, with an ensemble of SMBHBs as the source of GWB, this relation in the one-point function can be expected to manifest in the cumulants of the spectrum, as discussed in \cite{Lamb:2024gbh}. We also highlight that while the distribution of the mean is fairly consistent with a Gaussian distribution, the distribution of the skewness unequivocally deviates from this trend, consistent with our results based on simulating GWB maps and computing their one-point statistics. Our results show that the higher moments of the PTA residuals tend to a non-Gaussian ensemble distribution, attributed to a large scale quadrupolar correlation characteristic of a GWB. In particular, we find that the presence of large scale correlation implies that the central limit theorem cannot be applied. See Appendix \ref{sec:clt_and_large_scale_modes}.

However, we find that the dominant red noise component in the pulsars similarly produce a non-Gaussian ensemble distribution of sample statistics, although for a completely different reason compared with the GWB component. The variation of the red noise spectrum across pulsars is sourcing a huge departure of the resulting ensemble distribution away from a Gaussian distribution. This is notably more pronounced by orders of magnitude compared with the GWB, particularly for the higher moments of the pulsar timing residuals' one-point function. It is worth highlighting that the dominant red noise also completely contaminates the one-point statistics, as can be seen in the magnitudes of the variance and the kurtosis. When both contributions are considered in the simulations, among other usual components such as white noise \cite{Roebber:2019gha, NANOGrav:2020tig}, the resulting ensemble distributions do not turn away too far off from that of the red noise component alone. Adding more pulsars to the observation can be expected to only amplify the induced non-Gaussianity due to the variation of the red noise spectrum. Although one might expect the sum of uncorrelated random variables to approach a Gaussian distribution asymptotically, the noise is not identically distributed and different pulsars can exhibit orders of magnitude differences in their noise contributions. The cumulants of the red noise is dominated by a small number of pulsars with large noise components.  

\begin{figure}[h!]
    \centering
    \includegraphics[width=0.8\textwidth]{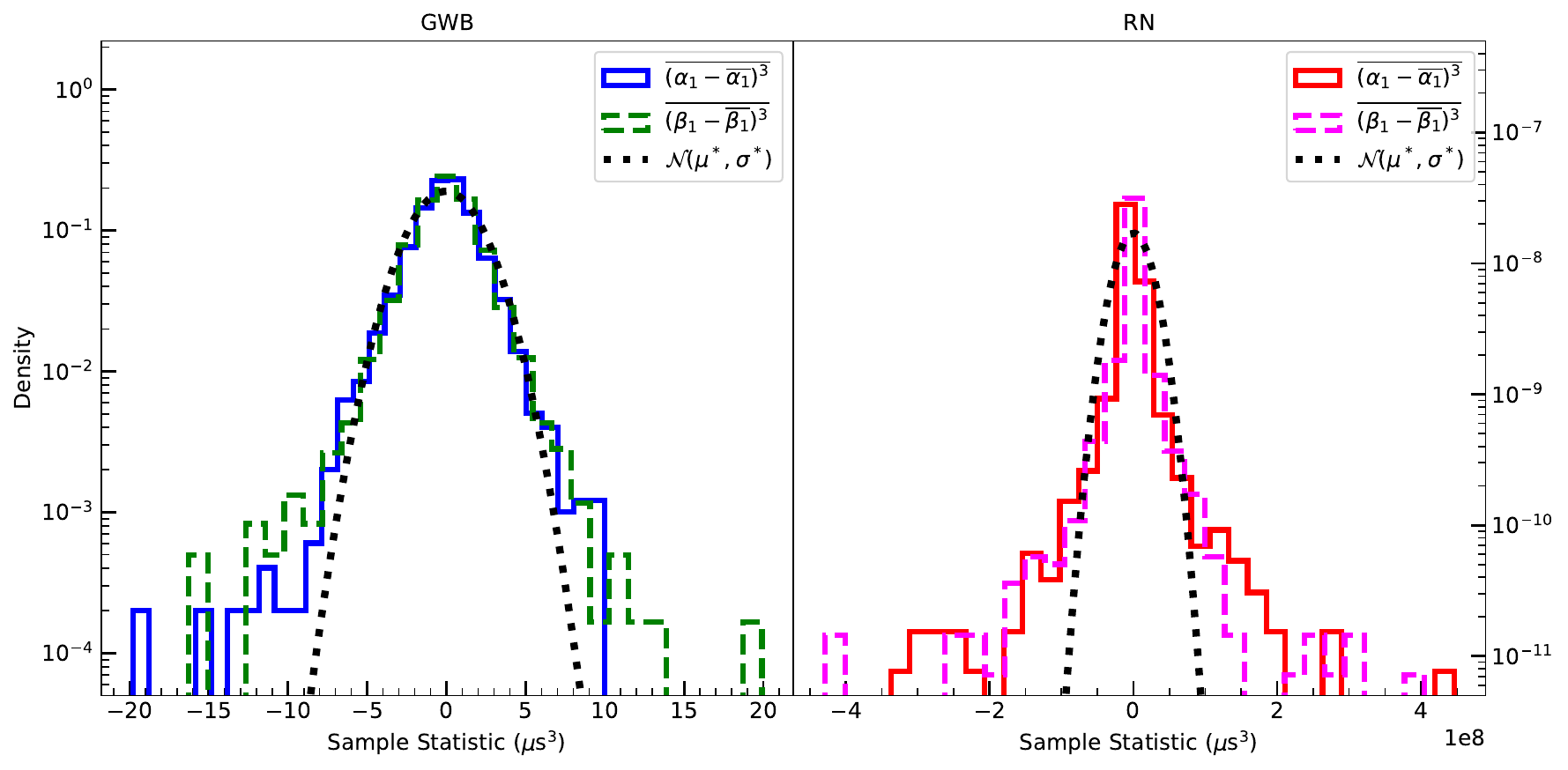}
    \includegraphics[width=0.8\textwidth]{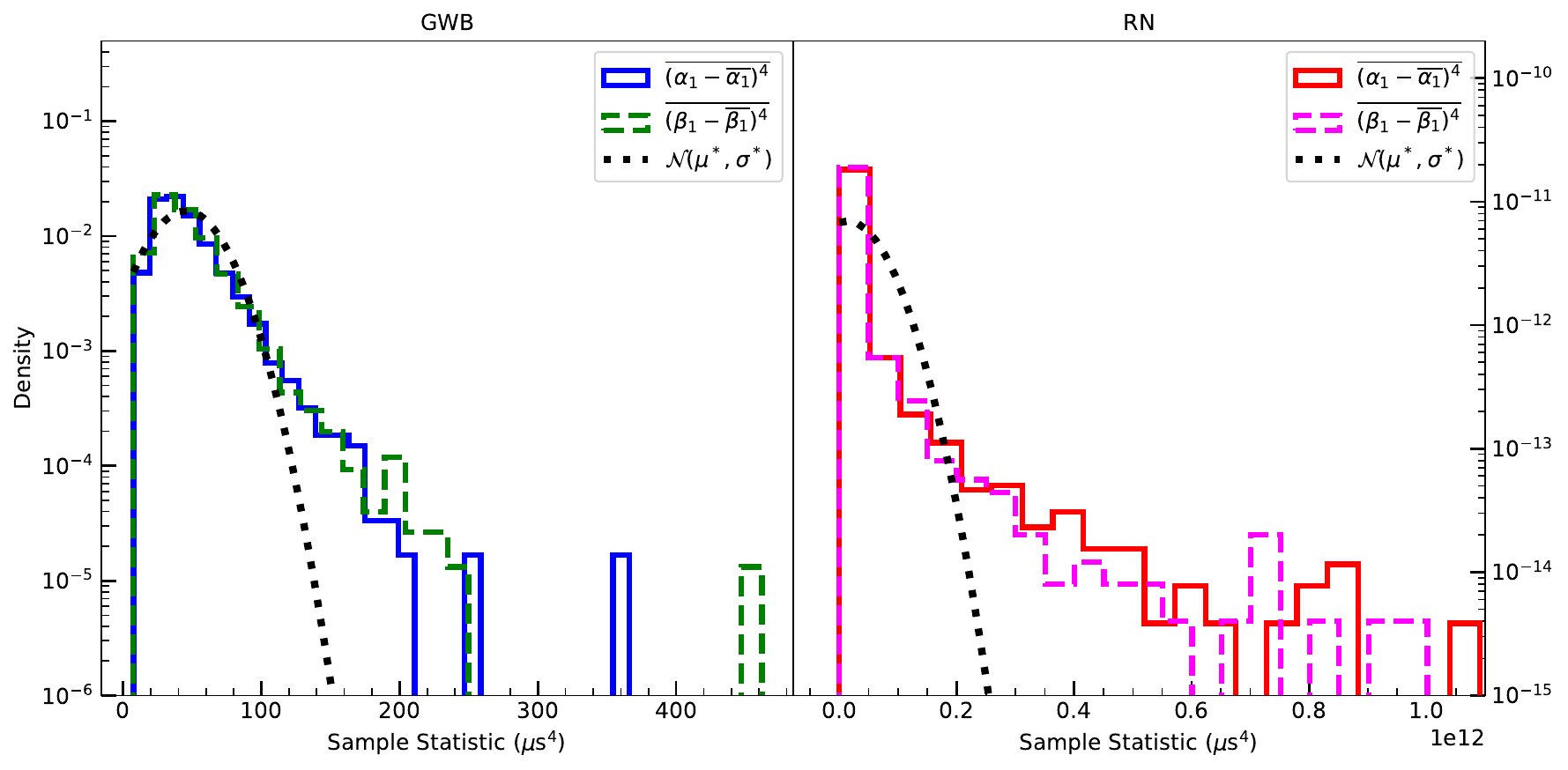}
    \caption{Distributions of the sample skewness, ${\cal S}_3$, and kurtosis, ${\cal K}_4$, in the first frequency bin for GWB and red noise (RN) components; $n_{\rm psrs}=100$ pulsars and $n_{\rm sims}=5000$ simulations; black dotted line corresponds to a Gaussian fit, ${\cal N}\left( \mu^*, \sigma^* \right)$, to the $\alpha_k$ distribution (added for visual purposes).}
    \label{fig:skewness_and_kurtosis_bin_1}
\end{figure}

This discussion is supported by zooming in on a single frequency bin. Figure \ref{fig:skewness_and_kurtosis_bin_1} shows the distribution of the skewness and the kurtosis in the first frequency bin in the simulated data, with roughly $f\sim 1$ nHz.

This illustrates that both the GWB and red noise components {produce a non-Gaussian} distribution {of the sample statistics across realisations} {; for GWB, this is due to the large scale correlation, whereas for the red noise, this is due to the variation of the spectrum across pulsars.} For the skewness, the ensemble distribution in both GWB and RN cases turns out to be narrower compared with a corresponding Gaussian one, but with a long tail. However, it must be emphasized that the sample statistic for red noise also varies by orders of magnitude compared with that of GWB. The same holds for the kurtosis, which manifests orders of magnitude difference in the red noise compared with GWB. This can be attributed to the variation in the red noise spectrum across pulsars, and can be tested by varying the number of pulsars. Note that the plots are shown in log-scale to make the differences perceivable by eye. To support of the above statements that were drawn from visual inspection, we also performed standard normality tests \cite{10.1093/biomet/58.2.341, 10.1093/biomet/60.3.613, 10.1093/biomet/52.3-4.591, PhipsonSmyth+2010, panagiotakos2008value, 2020SciPy-NMeth} that assign $p$-values relative to a null hypothesis that a sample comes from a normal/Gaussian distribution. For the variance, skewness, and kurtosis, in all frequency bins, the results have unambiguously ruled out the null hypothesis ($p \ll 0.05$), implying that the samples presented in Figure \ref{fig:ensemble_stats_all_bins}, hence also Figure \ref{fig:skewness_and_kurtosis_bin_1}, do not conform to a Gaussian ensemble distribution.

We emphasize that the notion of a Gaussian-distributed ensemble is distinct from the notion of Gaussianity of the field, which is the focus of this work. The underlying mechanism in both the GWB signal and the red noise are Gaussian processes. The results of this section highlight that GWB and red noise induce non-Gaussian distributions in their one-point function's sample statistics; in the case of GWB, this is due to the large scale correlation, while for red noise, this is due to the variation in the spectrum. It is worth noting too that the non-Gaussian ensemble distribution we find due to the dominating red noise in pulsars can be an artifact of our simulation. In the realistic case, there is only a small number of pulsars that can be identified as louder compared with the GWB.

However, our results indicate that the one-point function's statistics may be barely useful for determining a Gaussian GWB signal, because the red noise completely dominates the signal for PTAs. If the situation is reversed between signal and noise, such as expected in space-based GW detectors, then one-point statistics may be pursued to probe the Gaussianity of the GWB.

In this section, we have measured the one-point statistics of the GWB. In PTA, to our understanding, the signature of Gaussianity will be suppressed by noise in one-point function's cumulants. This brings us to pulsar-pair/two-point function statistics.

\section{Two-point statistics}
\label{sec:two_point_statistics}

In this section, we lay down our results on two-point function statistics of the GWB signal. We also consider the case when both signal and noise are added to the observation. {We add that the cumulants of the two-point function are operationally defined the same way as in the one-point case (third paragraph of Section \ref{subsec:the_gaussian_gwb}), except that the input are now a pair of residuals/Fourier bins, e.g., $\alpha_{ak} \beta_{bk}$ with $a \neq b$, and sample averages/statistics are gathered over pulsar pairs in a realisation.} 

The mean statistic, ${\cal E}_1$, of the timing residual cross correlation of a pulsar pair $a$ and $b$ due to an isotropic and Gaussian GWB \eqref{eq:gaussian_sgwb} can be shown to be
\begin{equation}
\label{eq:rarb_sgwb_mean}
    {\cal E}_1[ r_a r_b ] = \tilde{a}^2 c_{ab} \,,
\end{equation}
where $\tilde{a}$ is a constant related to the GWB spectrum, $c_{ab}= C\left( \hat{e}_a \cdot \hat{e}_b \right) + \delta_{ab}/2$, $C(x)$ is the HD curve, $C(x)=1/2 -y(x)/4 + 3 \left( y(x) \ln y(x) \right) / 2$, $y(x)=(1-x)/2$. Note that $c_{aa}=1$. Using Gaussian combinatorics, we can show that the variance, skewness and kurtosis of the timing residual correlation are given by
\begin{equation}
\label{eq:rarb_sgwb_variance}
    {\cal V}_2[r_a r_b] = \tilde{a}^4 \left( 1+c^{2}_{ab} \right) \,,
\end{equation}
\begin{equation}
\label{eq:rarb_sgwb_skewness}
    {\cal S}_3[r_a r_b] = 2 \tilde{a}^6 c_{ab} \left( 3 + c_{ab}^2 \right) \,,
\end{equation}
and
\begin{equation}
\label{eq:rarb_sgwb_kurtosis}
    {\cal K}_4[r_a r_b] = 3 \tilde{a}^8 \left( 3 + 14 c_{ab}^2 + 3 c_{ab}^4 \right) \,,
\end{equation}
respectively. Note that all of the above higher moments of the two-point function reduce to products of $c_{ab}$, by virtue of the Gaussianity of the underlying field. For this case, Gaussianity implies that the higher moments of the timing residual correlation give no more further information beyond the HD curve \cite{Planck:2019evm}. Since a GWB signal is differentiated from noise by spatial correlation, Gaussianity can be tested directly by measuring the higher moments of the two-point function, since only the signal can generate coherent correlations. 

\begin{figure}[h!]
    \centering
    \includegraphics[width=0.8\linewidth]{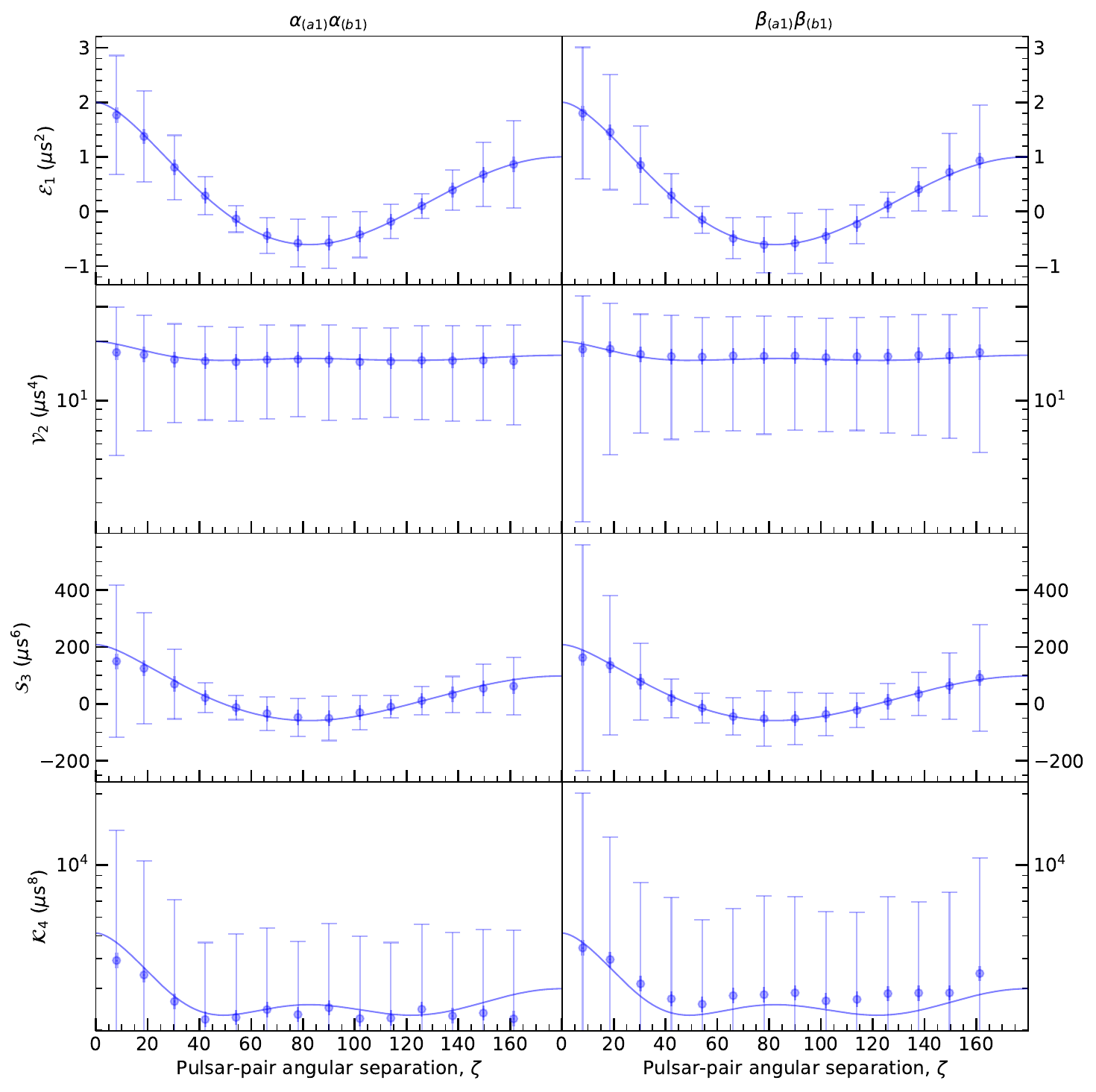}
    \caption{The signal of an isotropic, Gaussian nanohertz GWB in noise-free 30 yr-PTA simulations in the first frequency bin (100 pulsars, and 300 realizations). The theoretical curves are given by (\ref{eq:rarb_sgwb_mean}-\ref{eq:rarb_sgwb_kurtosis}). The results in the other frequency bins are similar, with relevant amplitudes scaled accordingly, $\tilde{a}^2 \rightarrow \tilde{a}^2 (f/f_1)^{-13/3}$, where $\tilde{a}^2$ and $f_1$ are the correlation amplitudes and the frequency in the first bin, respectively.}
    \label{fig:two_point_gwb_signal}
\end{figure}

To support the discussion, we simulate 30 yr-PTA realizations with one hundred pulsars, injected with a HD-correlated GWB with an amplitude $A_{\rm gw}=2.4 \times 10^{-15}$ and circular SMBHB spectral index $\gamma_{\rm gw}=13/3$ \cite{Phinney:2001di}. We pair up the pulsars in angular separation bins for every pair in a PTA realization, and then in each angular bin obtain the sample statistic (mean, variance, skewness, and kurtosis) of $\alpha_{ak} \alpha_{bk}$ and $\beta_{ak} \beta_{bk}$ over the pulsar pairs ($a,b$) and frequency bins, $f_k=k/T$. This is repeated $300$ times to create an ensemble, displaying the mean and the cosmic variance of the two-point function's sample statistics. {We add that in our analysis, all frequency and angular bins are treated independently. Optimal averages as in \cite{Allen:2022ksj, Allen:2024uqs} for two-point statistics will certainly be worth looking into for future work.}

% {We emphasize that this methodology is intuitive and can be afforded with simulations (providing a time series and the Fourier bins components for $n_{\rm psrs}$ pulsars), which has no significant gaps in the time-of-arrival, unlike real data. All we had to is loop over the frequency bins and the pulsar pairs, leading to a set $\{ \alpha_{a k} \beta_{b k} \}$ for $a \neq b$ (angular bin) and $k$ (frequency bin). The sample statistics are computed over the pulsar pairs in the set $\{ \alpha_{a k} \beta_{b k} \}$, which roughly has the same number of pulsars. The entire process from signal, and later noise, injection is repeated multiple times, with each simulation turning in a sample statistics representative of a probability distribution function (PDF) across realisations. In the noise-free case, the variance of this PDF is the cosmic variance, depicted in the following figures for the mean up to the kurtosis of the two-point function.}

The theoretical description (\ref{eq:rarb_sgwb_mean}-\ref{eq:rarb_sgwb_kurtosis}) is confirmed by our noise-free simulations, as in Figure \ref{fig:two_point_gwb_signal}, obtained by injecting only the GWB signal into the pulsars. The results also confirm that the signal in the higher frequency bins are described simply by the same signal in the first bin, with the amplitudes rescaled accordingly to the power spectrum of the signal, $\tilde{a}^2 \rightarrow \tilde{a}^2 (f/f_1)^{-13/3}$, where $\tilde{a}^2$ and $f_1$ are the correlation amplitudes and the frequency in the first bin, respectively. This shows that the higher moments of a Gaussian GWB's two-point function in a PTA can be described by the spectrum and spatial correlation that is already associated with its mean.

\begin{figure}[h!]
    \centering
    \includegraphics[width=0.8\linewidth]{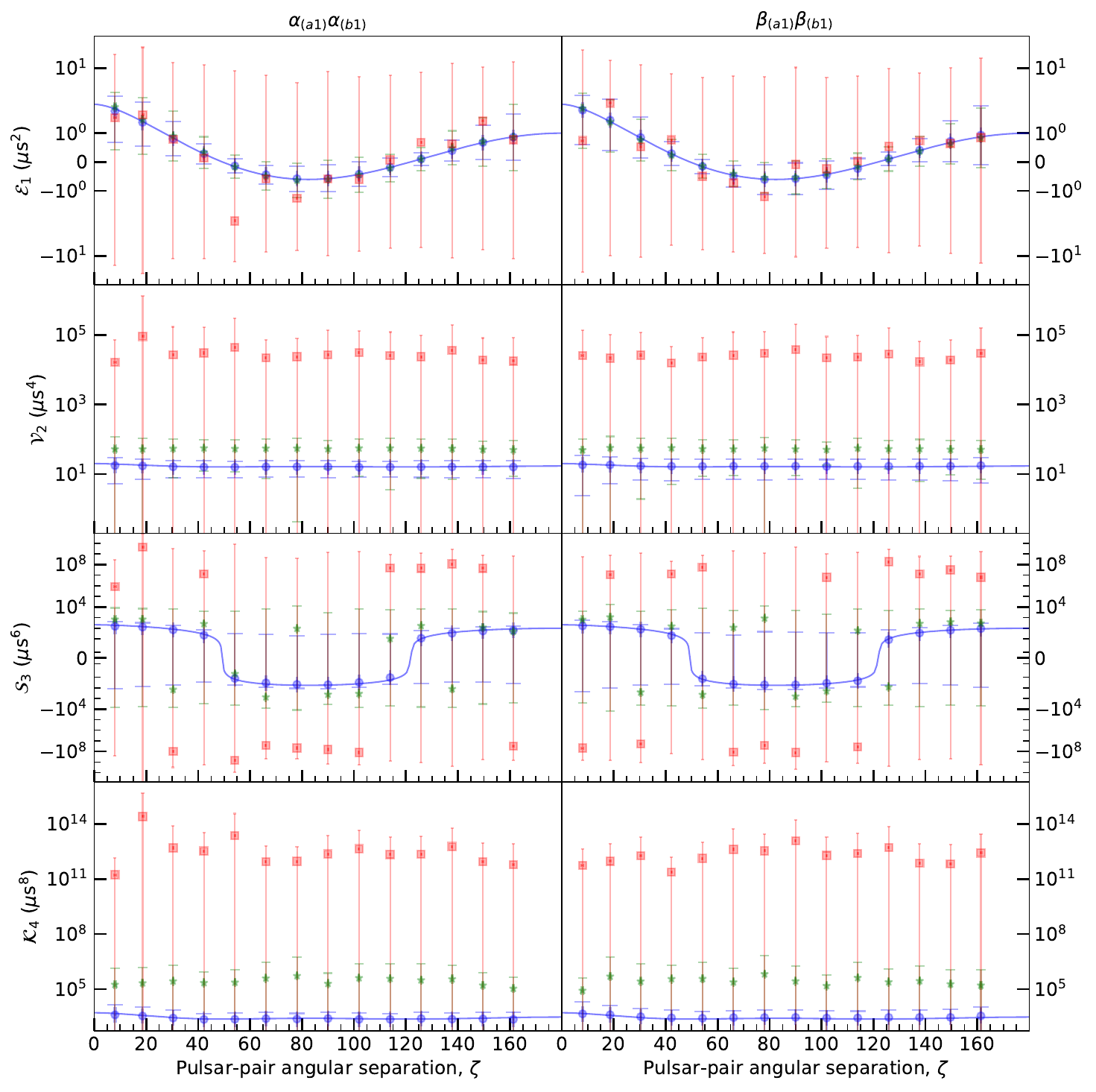}
    \caption{The timing residual correlation two-point statistics in 30 yr-PTA simulations in the first frequency bin (100 pulsars, and 300 realizations) with varying degrees of noise in the pulsars. Blue is noise-free (GWB signal); green is with soft noise (RN as loud as signal); red is with mild noise (RN one order of magnitude above signal). The theoretical curves (blue) are given by (\ref{eq:rarb_sgwb_mean}-\ref{eq:rarb_sgwb_kurtosis}).}
    \label{fig:two_point_pta_signal_and_noise}
\end{figure}

\begin{figure}[h!]
    \centering
    \includegraphics[width=0.8\linewidth]{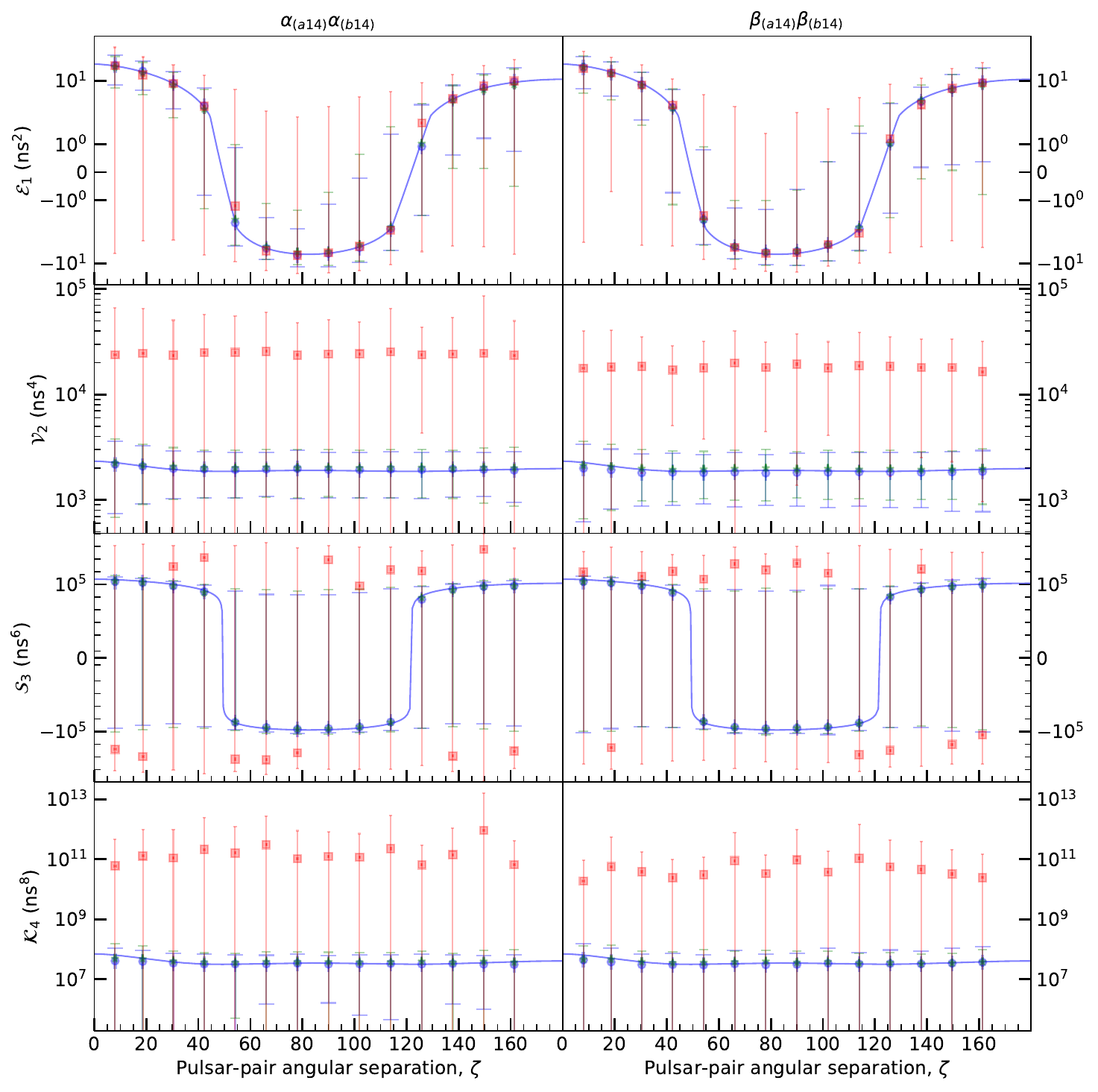}
    \caption{The timing residual correlation two-point statistics in 30 yr-PTA simulations (same as in Figure \ref{fig:two_point_pta_signal_and_noise}) in the 14th frequency bin (100 pulsars, 300 realizations) with varying degrees of noise in the pulsars. Blue is noise-free (GWB signal); green is with soft noise (RN as loud as signal); red is with mild noise (RN one order of magnitude above signal). The theoretical curves (blue) are given by (\ref{eq:rarb_sgwb_mean}-\ref{eq:rarb_sgwb_kurtosis}), with the amplitude rescaled as $\tilde{a}^2 k^{-13/3}$.}
    \label{fig:two_point_pta_signal_and_noise_f14}
\end{figure}

In the astrophysical PTA setting the signal is dominated by red noise in pulsars that has similar time scales as GWB timing residuals. This renders the one-point statistics unusable for testing the Gaussianity of the signal {(Section \ref{subsec:one_point_statistics})}. To understand how this manifests in the two-point function, we can take an observation{/timing residual} from a pulsar $a$ to be $r_a = s_a + n_a$ where $s_a$ is the GWB signal (satisfying (\ref{eq:rarb_sgwb_mean}-\ref{eq:rarb_sgwb_kurtosis})) and $n_a$ is the noise. In a pure-noise case, the analogous two-point function statistics can be obtained simply by replacing the HD correlation, $c_{ab}$, in (\ref{eq:rarb_sgwb_mean}-\ref{eq:rarb_sgwb_kurtosis}) by $\delta_{ab}$. Our pure-noise simulations, opposite to the noise-free case, support this simple model of an uncorrelated noise in the higher moments of the two-point function. However, in the realistic scenario where both signal and noise are present in the observation, a more precise way to model an (effective) correlation is by expanding the powers of $r_a r_b \equiv (s_a + n_a) (s_b + n_b)$, subject to simplifications that apply such as Gaussian factorization. Continuing it this way is tedious, and can perhaps pay off as a way to include the information in the two-point function for GWB detection purposes. We shall leave this for future work. Nonetheless, our simulations can tease the features to expect in the two-point function when noise is present as well as the signal, as shown in Figures \ref{fig:two_point_pta_signal_and_noise} and \ref{fig:two_point_pta_signal_and_noise_f14} in the 1st ($f_1\sim 1.06$ nHz) and 14th ($f_{14}\sim 14.8$ nHz) frequency bins of the data.

In the analysis, we consider several levels of the noise relative to the amplitude of the GWB signal ($A_{\rm gw}=2.4\times 10^{-15}$). First is the noise-free level, for which case the ensemble uncertainty can be identified with the cosmic variance \cite{Roebber:2015iva, Roebber:2016jzl, Allen:2022dzg, Bernardo:2022xzl}. Next one is `soft' noise, when the red noise spectrum in the pulsars in each realization is drawn with amplitudes $\log_{10}A_{\rm rn}={\cal U}(-17, -15)$ and spectral shape indices $\gamma_{\rm rn}={\cal U}(2, 6)$, resulting to the overall noise becoming only as loud as the signal in the observation. We also consider a `mild' noise case when instead the noise can be drawn up to an order of magnitude above the signal, $\log_{10}A_{\rm rn}={\cal U}(-17, -14)$. Figures \ref{fig:two_point_pta_signal_and_noise}-\ref{fig:two_point_pta_signal_and_noise_f14} illustrate the resulting ensemble statistics of the higher moments of the timing residual correlation for the noise-free, soft-, and mild-noise scenarios. {Note that the same signal in Figure \ref{fig:two_point_gwb_signal} is shown in Figures \ref{fig:two_point_pta_signal_and_noise}-\ref{fig:two_point_pta_signal_and_noise_f14} (blue points traced by the theoretical (\ref{eq:rarb_sgwb_mean}-\ref{eq:rarb_sgwb_kurtosis})), except with the mean and the skewness expressed in log-scale for visual clarity when the noise is taken into account.}

The results show that despite being spatially-uncorrelated and drawn from independent Gaussian random processes the red noise remains to dominate the moments of the two-point function in the first frequency bin, particularly, in the variance and the kurtosis, even in the soft and mild PTA noise scenarios we considered. This domination of the noise is only amplified in the loud noise case. Nonetheless, the figures suggest that even with only one hundred pulsars (consistent with present PTA data) the mean of the two-point function can already hint toward a detection of the signal, which may be followed up by searching for the signal of Gaussianity (\ref{eq:rarb_sgwb_mean}-\ref{eq:rarb_sgwb_kurtosis}) in the higher moments. This is reminiscent of the present state-of-the-art PTA science, indicating a compelling evidence of the sought GWB signal. 

We observe that the difference in orders of magnitude between signal and noise can drop significantly in the higher frequency bins, which may eventually help when searching for the signal of Gaussianity through the two-point function in present and future PTA data. This is illustrated in the 14th frequency bin shown in Figure \ref{fig:two_point_pta_signal_and_noise_f14}. The theoretical curve is drawn with the same constant fitted in the first bin, $\tilde{a}^2$, only rescaled according to the expected power spectrum of the signal, $\tilde{a}^2 \rightarrow \tilde{a}^2 (f_{14}/f_1)^{-13/3}$. Note the units of the correlation/two-point function of ns$^2$ (nanosecond squared) in Figure \ref{fig:two_point_pta_signal_and_noise_f14} compared with $\mu$s$^2$ (microsecond squared) in Figure \ref{fig:two_point_pta_signal_and_noise}. The 14th bin thus tells nothing about the signal that is not already known in the other bins of the data. However, the frequency dependence may turn out to play a bigger role when noise is considered, which is after all most important for detection purposes, as shown in Figure \ref{fig:two_point_pta_signal_and_noise_f14} for the soft and mild noise cases. We see that the signal has overcome soft noise in the 14th bin, and is dominant over it in the 1st bin. Another way we can put this is that the same noisy two-point function data in the 1st frequency bin nearly coincides with the predicted signal in the 14th frequency bin. Understandably, the noise continues to be dominant compared to the signal in the mild noise case. Nonetheless, there is now a clearer evidence of the signal in the mean of the two-point function in the 14th bin compared with the 1st bin. The difference in orders of magnitude in the higher cumulants of the two-point function has also significantly dropped.

We have confirmed the same conclusions on two-point function statistics with 200 and 400 pulsars. {For these cases, the significance of the signal with respect to the noise in the two-point statistics has increased as expected, e.g., compared to 100 pulsars, in our mild noise scenario, the variance of the noise compared to the signal in the 14th frequency bin has decreased by 4\% and 10\% for 200 and 400 pulsars, respectively. Accurately modelling the noise (as described in the 5th paragraph of this section) should further enable a more precise determination of the signal's properties via two-point statistics.} In this work we focus on simulations with 100 pulsars because that is the current state-of-the-art in the field. {The conclusions on one- and two-point statistics are also robust to the consideration of pulsar white noise, which is uncorrelated spatially and in time, and has negligible contribution. We leave for future work the analysis with dispersion measure/chromatic noises which depend on frequency of radio observation and may be relevant for some pulsars in a PTA.}

\section{Outlook}
\label{sec:outlook}

In this work, we have studied the signatures of an isotropic and Gaussian GWB in the one- and two-point statistics, setting a baseline for future analysis that may go toward a detection of not only a GWB but also of confirmation or repudiation of its Gaussian nature. This adds another layer of potential evidence to consider in the debated source of the common spatially-correlated signal in PTAs, since astrophysical and cosmological sources of GWB have differing levels of Gaussianity that can be expected to manifest observationally. Our results have shown that one-point statistics will be completely dominated by noise. Nonetheless, our two-point statistical analysis hints that the signatures of a Gaussian GWB may eventually manifest in the higher cumulants of the two-point function in PTA data.

Our analysis has no assumption about the source statistics, but it will be important to draw this connection at some point {\cite{Allen:2022dzg, Allen:2024rqk}}. It remains to apply our tests to present data and forecast results expected in the forthcoming PTA/SKA precision era \cite{Lazio:2013mea, Weltman:2018zrl, ChandraJoshi:2022etw, Caliskan:2023cqm}, which we will defer to future work. A Gaussian signal may turn out to be more significant in real data than is implied here, even though we suspect that the non-Gaussianity due to the noise will continue to dominate the cumulants. This is because the intrinsic noise parameters of each pulsar are measured, to an extent, post-sampling in the standard PTA search pipeline for a common spatially-correlated signal. The constrained noise parameters, particularly of the louder pulsars, may eventually be used to reduce their influence on the cumulants in both one- and two-point statistics. Real data is of course trickier, and there are plenty of challenges with PTA data that must be dealt with such as the practical limits to the sensitivity curve \cite{Hazboun:2019vhv}, and the fact that the timing data contains gaps. 

On the other hand, the statistical properties of a Gaussian GWB in the one- and two-point function discussed in this work are general and independent of any particular primordial model. The non-Gaussianity in the one-point cumulants (left of Figure \ref{fig:skewness_and_kurtosis_bin_1}) and the two-point signal (\ref{eq:rarb_sgwb_mean}-\ref{eq:rarb_sgwb_kurtosis}) are tied to the quadrupolar spatial correlation, due to the tensorial nature/polarization of GWs. This implies that analogous tests may be setup to study GWB in other GW bands, such as for future space based GW detectors that are expecting to meet a foreground GWB from white dwarfs, among other relevant cosmological GWBs motivated in theory. The formalisms drawn out in this work are easily extendible to accommodate non-Einsteinian and subluminal GW propagations \cite{Chamberlin:2011ev, Qin:2020hfy, NANOGrav:2021ini, Chen:2021wdo, Chen:2021ncc, Wu:2023pbt, Bernardo:2022vlj, Bernardo:2022rif, Bernardo:2023mxc, Bernardo:2023pwt, Liang:2023ary, Cordes:2024oem}. When the tests come to fruition, it will be inevitably important to relate constraints on non-Gaussianity in the GWB observables to theory, requiring predictions beyond the power spectrum \cite{Powell:2019kid, Tasinato:2022xyq, Zhu:2022bwf}. We leave this to future work.

\section*{Data availability}
Python notebooks that reproduce our results are published in \href{https://github.com/reggiebernardo/notebooks/tree/main/supp_ntbks_arxiv.2407.17987}{GitHub}.

\acknowledgments
The authors thank Achamveedu Gopakumar, Subhajit Dandapat, and Debabrata Deb for numerous fruitful discussions and William Lamb for important comments on a preliminary draft. RCB and SA are supported by an appointment to the JRG Program at the APCTP through the Science and Technology Promotion Fund and Lottery Fund of the Korean Government, and was also supported by the Korean Local Governments in Gyeongsangbuk-do Province and Pohang City. This work was supported in part by the National Science and Technology Council of Taiwan, Republic of China, under Grant No. NSTC 113-2112-M-001-033. 

% \paragraph{Note added.} This is also a good position for notes added
% after the paper has been written.

\appendix
% \section{Some title}
% Please always give a title also for appendices.

\section{Central Limit Theorem and Large Scale Modes}
\label{sec:clt_and_large_scale_modes}

The central limit theorem and law of large numbers play an important role in cosmology \cite{Verde:2009tu, Trotta:2017wnx}. A sample (spatial) average of a summary statistic, constructed from a cosmological data set, is assumed to converge to the true mean of the underlying probability distribution of that statistic. Beyond that, we also often assume that the covariance associated with said measurement is Gaussian. For example, if we measure some summary statistic $u_{\mu}$\footnote{This could be the two-point correlation function, the Minkowski Functionals etc. We include a $\mu$ subscript to denote a possible binning of the statistic in density, separation etc. In this appendix we simply measure the cumulants of a field, which are scalars.} then the sample average is taken to be an unbiased estimator of the ensemble average $\langle u_{\mu}\rangle$, and the uncertainty of the measurement is inferred from a covariance matrix $\Sigma_{\mu\nu}$. This can be constructed either by numerical estimation from mock data or explicitly calculating the ensemble average $\Sigma_{\mu\nu} \sim \langle u_{\mu} u_{\nu} \rangle$. By using the ensemble average $\langle u_{\mu}\rangle$ and covariance $\Sigma_{\mu\nu}$ to quantify the statistical properties of $u_{\mu}$, we are effectively modelling it as a multivariate Gaussian. In this appendix, we will consider the extent to which these approximations can be made. 

We start with an all-sky map of a Gaussian random field $\Delta_{i}$ drawn from an angular power spectrum $C_{\ell}$, where $i$ denotes the pixel number on the two-sphere\footnote{To generate and manipulate the maps we utilise Healpix (http://healpix.sourceforge.net) \citep{Gorski:2004by}.}. We measure the cumulants of this field, because they constitute the simplest set of non-trivial summary statistics that can be used to present our point. We define 
\begin{equation}\label{eq:kapn} \kappa_{n} \equiv   {1 \over N_{\rm pix}} \sum_{i=1}^{N_{\rm pix}} \Delta_{i}^{n} .
\end{equation}
For an individual pixel, we write down the probability distribution function of $\Delta_{i}^{n}$ as 

\begin{equation} P_{\Delta_{i}^{n}}(x) = {|x|^{(1-n)/n} \over \sqrt{2\pi\sigma^{2}}} e^{-|x|^{2/n}/2\sigma^{2}} ,
\end{equation}

\noindent where $x \in \mathbb{R}_{> 0}$ for $n$ even and $x \in \mathbb{R}$ for $n$ odd. Each pixel $\Delta_{i}$ is correlated according to the covariance matrix $\Sigma_{ij} = \zeta(\theta_{ij})$, where $\zeta(\theta_{ij})$ is the angular correlation function and $\theta_{ij}$ is the angular separation of pixels $i$ and $j$. 

The first cumulant, $\kappa_{1}$ is itself a Gaussian random variable. If the pixels are uncorrelated and $\Sigma_{ij} = \sigma^{2} \delta_{ij}$, then $\kappa_{1} \sim {\cal N}(0,\sigma^{2}/N_{\rm pix})$. In the presence of correlations, we have $\kappa_{1} \sim {\cal N}(0,\zeta/N^{2}_{\rm pix})$, where $\zeta = \sum_{i,j} \Sigma_{ij}$. Because we mean subtract the field, we do not consider $\kappa_{1}$ further. 

The second cumulant $\kappa_{2}$ is a non-Gaussian random variable. By performing a spectral decomposition of the covariance matrix -- $\Sigma = U^{\rm T} \Lambda U$, where $U U^{\rm T} = I$ and $\Lambda_{ij} = \lambda_{i}\delta_{ij}$, we can write 
\begin{equation}\label{eq:kap2_de} \kappa_{2} =  {1 \over N_{\rm pix}}\sum_{i=1}^{N_{\rm pix}} \Delta_{i}^{2} = {1 \over N_{\rm pix}}\sum_{i=1}^{N_{\rm pix}} \lambda_{i} y_{i}
\end{equation} 
where $y_{i}$ are independent, $\chi^{2}(1)$ random variables and $\lambda_{i}$ are the eigenvalues of the pixel covariance matrix $\Sigma$. We have defined $\chi^{2}(n)$ as a chi square distribution with $n$ degrees of freedom. The probability distribution function (PDF) of $\kappa_{2}$ does not have a closed form solution for the general case in which $\lambda_{i} \neq \lambda_{j}$ for $i \neq j$, but it is straightforward to extract its moments from the second sum in equation (\ref{eq:kap2_de}) since $y_{i}$ are independent, identically distributed (iid). For example the mean and variance of $\kappa_{2}$ are given by
\begin{eqnarray} 
\label{eq:ekap2} {\rm E}(\kappa_{2}) &=& {1 \over N_{\rm pix}} \sum_{i=1}^{N_{\rm pix}} \lambda_{i} {\rm E}(y_{i}) = {1 \over N_{\rm pix}} \sum_{i=1}^{N_{\rm pix}} \lambda_{i} \\
\label{eq:vkap2} {\rm var}(\kappa_{2}) &=&  \sum_{i=1}^{N_{\rm pix}} {\rm var} \left({\lambda_{i} \over N_{\rm pix}} y_{i} \right)  = {2 \over N_{\rm pix}^{2}} \sum_{i=1}^{N_{\rm pix}} \lambda^{2}_{i} \,.
\end{eqnarray} 

For the hypothetical case in which all pixels are uncorrelated, we have $\Sigma_{ij} = \sigma^{2}\delta_{ij}$, $\lambda_{i} = \sigma^{2}$ and 
\begin{equation}\label{eq:kap2_sa} \kappa_{2} =  {\sigma^{2} \over N_{\rm pix}}\sum_{i=1}^{N_{\rm pix}}  y_{i}  = {\sigma^{2} \over N_{\rm pix}} \chi^{2}(N_{\rm pix})
\end{equation} 
and hence $\kappa_{2}$ follows a $\chi^{2}(N_{\rm pix})$ distribution, with $N_{\rm pix}$ degrees of freedom. In the limit $N_{\rm pix} \to \infty$, we can informally write $\chi^{2}(N_{\rm pix}) \sim {\cal N} (N_{\rm pix}, 2N_{\rm pix})$, as expected from the central limit theorem. 

\begin{figure}[t]
    \centering
  \includegraphics[width=0.45\textwidth]{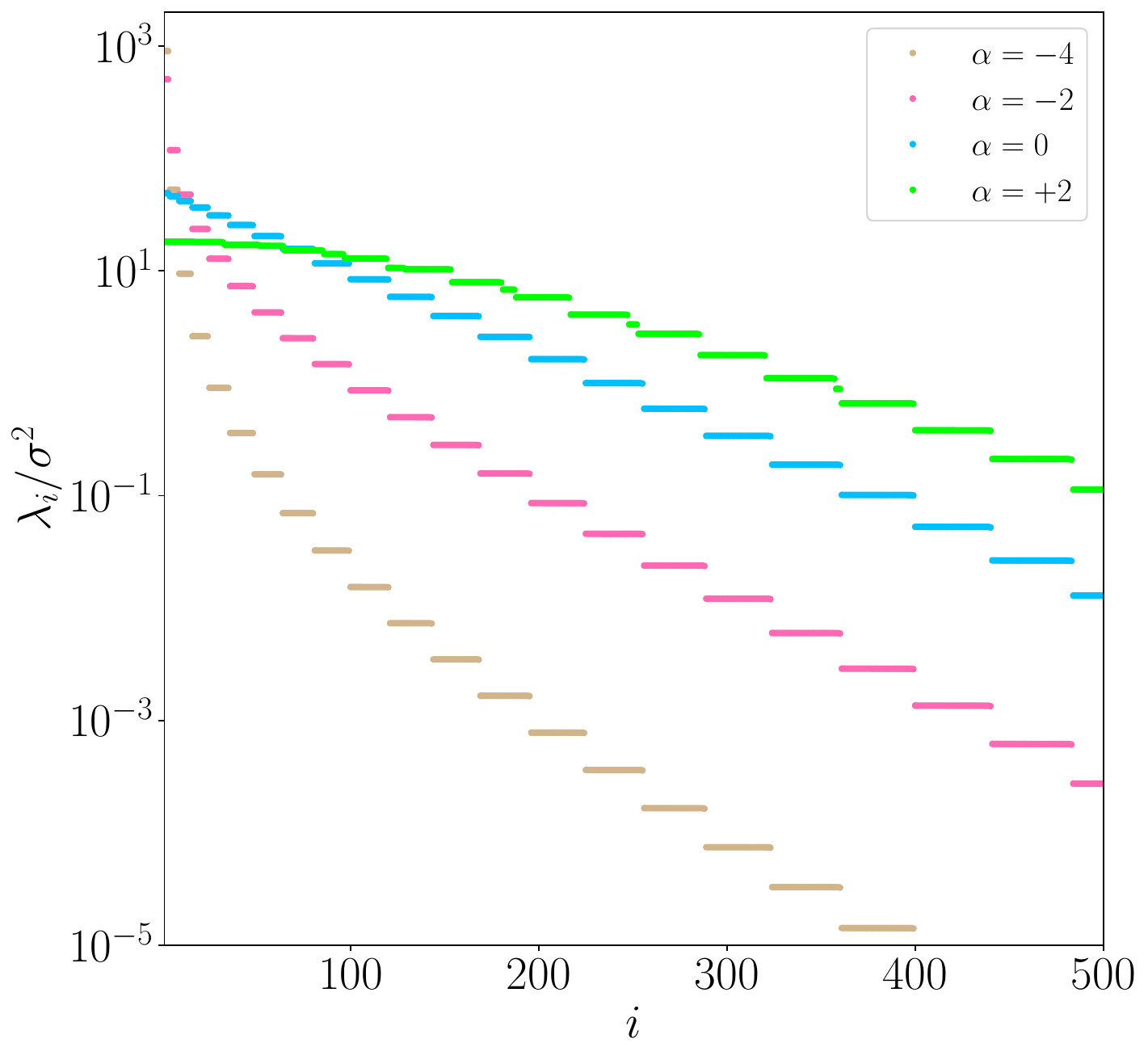}
    \caption{Eigenvalues of pixel covariance matrix $\Sigma_{ij}$ for Gaussian random fields on the sphere generated with different angular power spectra $C_{\ell} = \ell^{\alpha}$.}
    \label{fig:cor_lam}
\end{figure}

However, the presence of correlations between pixels changes the probability distribution of $\kappa_{2}$. The eigenvalues $\lambda_{i}$ become increasingly hierarchical as large scale correlation is introduced to the field $\Delta_{i}$. To show this, we adopt a power law angular power spectrum $C_{\ell} = \ell^{\alpha}$ and generate Gaussian random fields $\Delta$ on the two-sphere. We generate a low resolution pixel map $N_{\rm pix} = 12\times n_{\rm side}^{2}$ with $n_{\rm side} = 16$, and construct the covariance matrix between the pixels according to 
\begin{equation}\label{eq:zeti} \Sigma_{ij} = \sum_{\ell} {2\ell+1 \over 4\pi} C_{\ell}W_{\ell}^{2}(\theta_{G}) P_{\ell}(x_{ij}) \end{equation} 
where we smooth the field with a Gaussian kernel $W_{\ell} = \exp[-\ell (\ell+1)\theta_{G}^{2}/2]$ and angular smoothing scale  $\theta_{G} = 2 \sqrt{4\pi/N_{\rm pix}}$. $P_{\ell}(x)$ are the Legendre polynomials and $x_{ij} = \hat{n}_{i} . \hat{n}_{j}$, where $\hat{n}_{i}$ is the unit vector pointing to pixel $i$\footnote{In equation (\ref{eq:zeti}) have used the fact that the data is an all-sky map, for point distributions or masked data a different estimator for $\Sigma_{ij}$ should be used.}. The diagonal elements of $\Sigma_{ij}$ are 
\begin{equation}  \sigma^{2} \equiv \Sigma_{ii}  = \sum_{\ell} {2\ell+1 \over 4\pi} C_{\ell}W_{\ell}^{2}(\theta_{G}) \,. \end{equation}

Once we have the covariance matrix we decompose it according to $\Sigma = U^{\rm T} \Lambda U$. In Figure \ref{fig:cor_lam} we present the eigenvalues; $\Lambda_{ij} = \lambda_{i}\delta_{ij}$ of the $\Sigma_{ij}$ covariance matrix for different $\alpha$ power law spectra. For visual clarity we normalise $\lambda_{i}$ by $\sigma^{2}$.

\begin{figure}[t]
    \centering
  \includegraphics[width=0.8\textwidth]{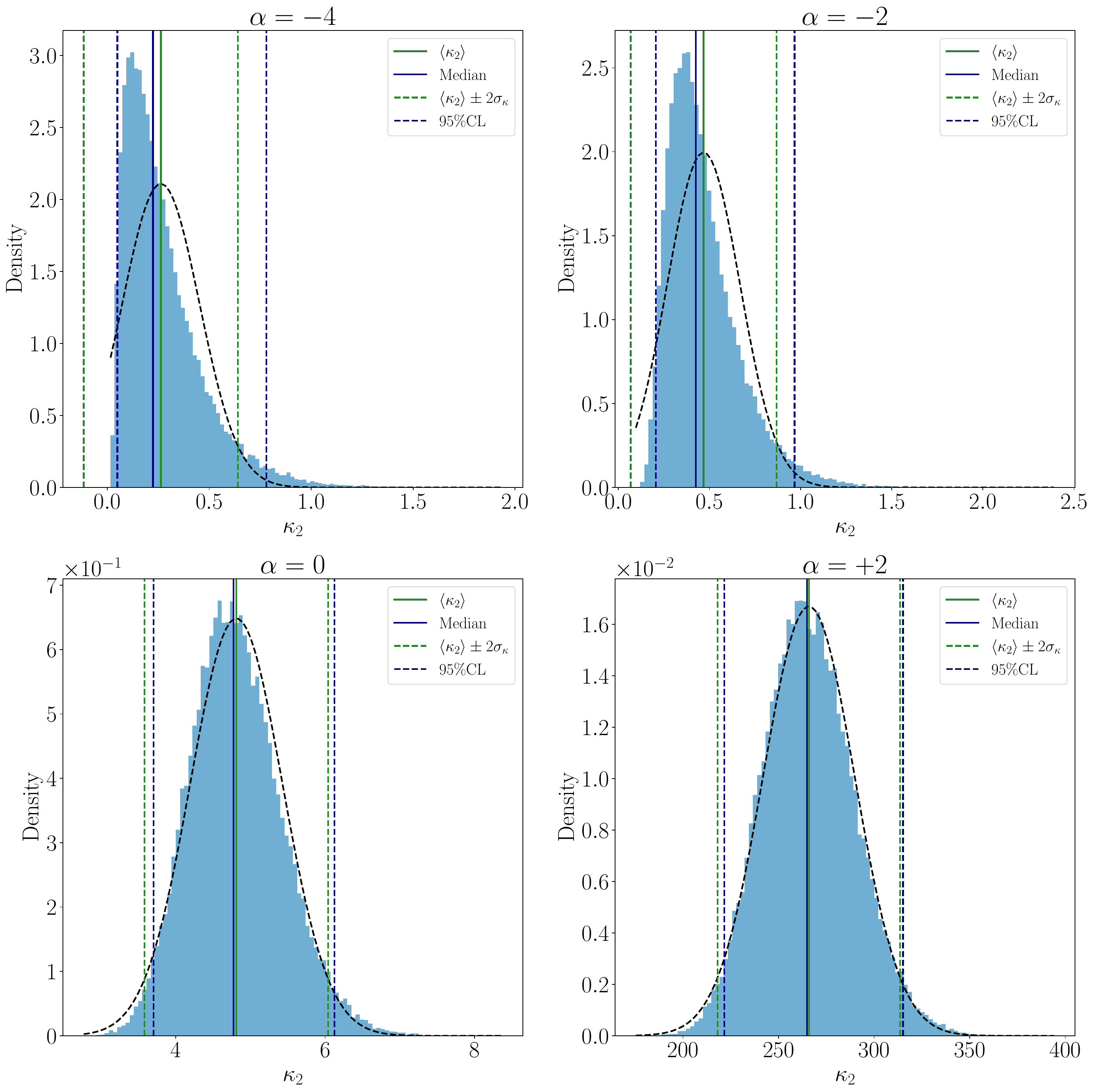}
    \caption{Numerically reconstructed PDFs (blue histograms) of the cumulant $\kappa_{2}$ from all-sky distributions of Gaussian random fields, generated with angular power spectrum $C_{\ell} = \ell^{\alpha}$, with $\alpha = -4,-2,0,2$ (top left, top right, bottom left, bottom right). The black dashed lines are Gaussian approximations of the distribution, using the expectation value $\langle \kappa_{2}\rangle$ and covariance $\langle \kappa_{2}^{2} \rangle$. }
    \label{fig:p2}
\end{figure}

We observe that as the power spectrum $C_{\ell}$ becomes increasingly blue tilted, the eigenvalues become sharply peaked at $i\sim 1$. Because the eigenvalues act as coefficients multiplying the random variables $y_{i}$ in the expression (\ref{eq:kap2_de}) for $\kappa_{2}$, the presence of a hierarchy implies that the sum will be dominated by a small number of random variables. In this case, the large $N_{\rm pix}$ limit will not Gaussianize the $\kappa_{2}$ PDF. Indeed, as the hierarchy in $\lambda_{i}$ becomes increasingly pronounced, $\kappa_{2}$ will be approximated as being drawn from a $\chi^{2}(n)$ distribution with $n \sim {\cal O}(1)$. In Figure \ref{fig:p2} we present the PDF of $\kappa_{2}$ for $\alpha = -4, -2, 0, 2$. The blue histograms were obtained by generating $N_{\rm real} = 10^{4}$ realisations of a Gaussian random field on the sphere for each $C_{\ell} = \ell^{\alpha}$ and estimating $\kappa_{2}$ by taking the pixel sum (\ref{eq:kap2_de}). The black dashed lines in Figure \ref{fig:p2} are Gaussian distributions with mean and variance given by equations (\ref{eq:ekap2},\ref{eq:vkap2}) and the vertical green solid/dashed lines are the mean and $\pm 2\sigma$ bounds of this Gaussian approximation. The blue vertical lines are the median and $95\%$ confidence region of the numerically reconstructed PDF.

We clearly observe that the approximate Gaussianity of the summary statistic is strongly model dependent, and the numerically reconstructed PDF is Gaussianized as we move from a blue tilted $\alpha = -4$ (top left panel) to red tilted $\alpha = +2$ (bottom right panel) power spectrum. When the data exhibits large scale power, we cannot expect the cumulants to follow a Gaussian ensemble distribution. 

\begin{figure}[h!]
    \centering
  \includegraphics[width=0.8\textwidth]{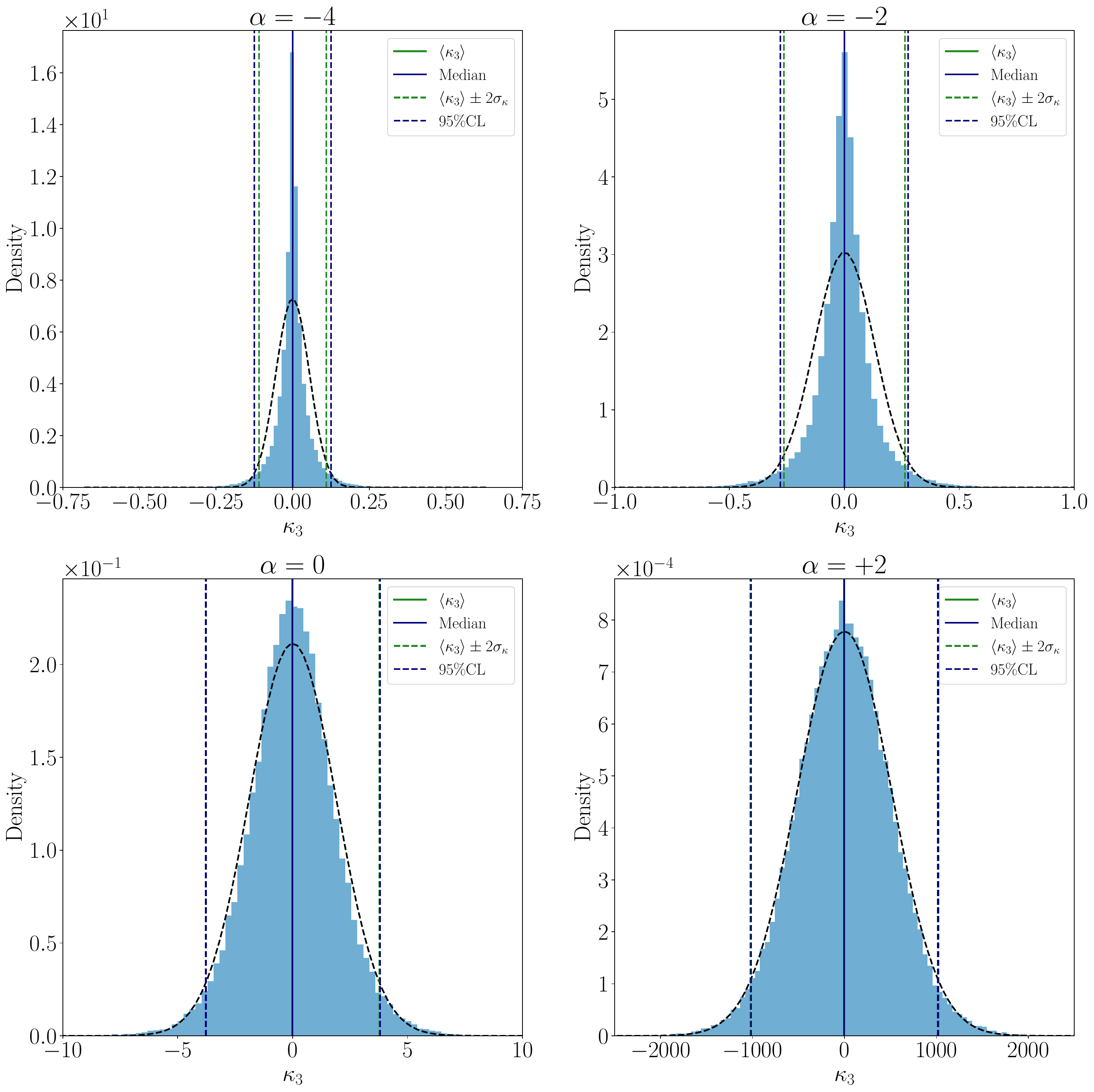}
    \caption{Same as Figure \ref{fig:p2}, but for the cubic cumulant $\kappa_{3}$. }
    \label{fig:s3}
\end{figure}

\begin{figure}[h!]
    \centering
    \includegraphics[width=0.4\textwidth]{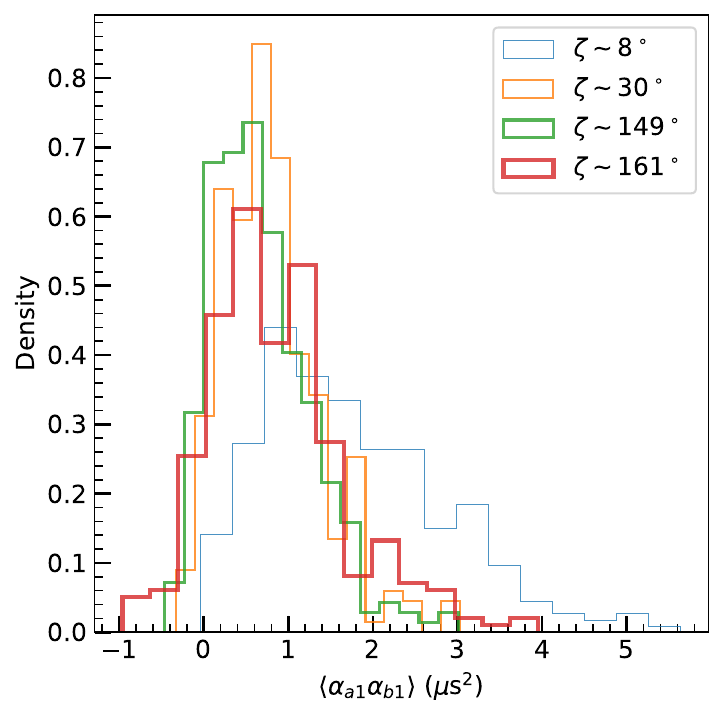}
    \includegraphics[width=0.4\textwidth]{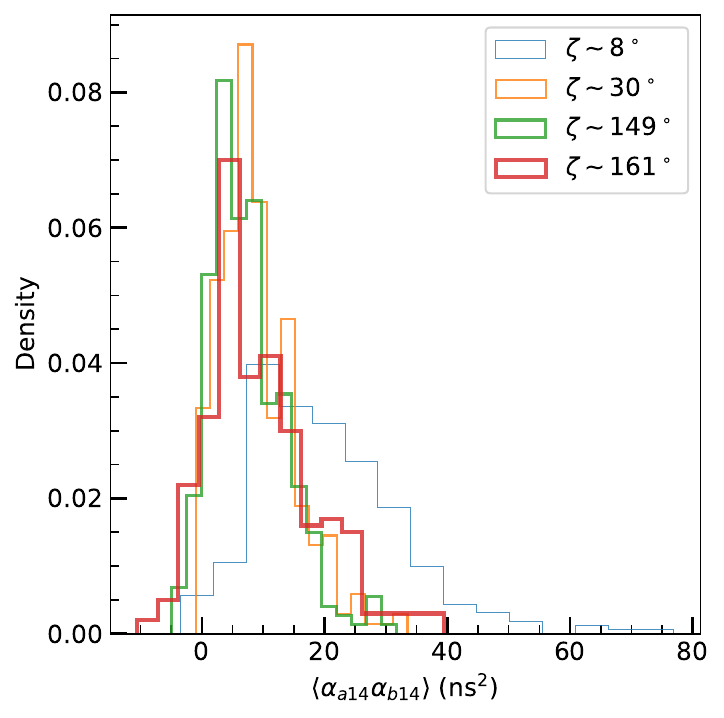}
    \caption{Ensemble distributions of the two-point function's sample mean (in the 1st and 14th frequency bins) at fixed angular bins $\zeta$ for noise-free PTA simulations with 100 pulsars and 300 realisations. Notice the ns$^2$ units in the 14th bin.}
    \label{fig:ensemble_ngwb_2pfs}
\end{figure}

The higher order cumulants are less tractable than the variance, but the same logic applies. The skewness $\kappa_{3}$ involves triple products of the Gaussian random variable $\Delta_{i}$, which are less straightforward to de-correlate than the variance. However, the same issues arise -- in the presence of large scale correlations in the data the probability distributions of the summary statistic is highly non-Gaussian. In Figure \ref{fig:s3} we present the PDFs of $\kappa_{3}$ for $\alpha = -4, -2, 0, +2$. As before, the solid/dashed blue lines are the median and $95\%$ confidence region of the numerically reconstructed PDF (cf blue histograms). The black dashed curve is a Gaussian distribution inferred from the ensemble average of the statistics \cite{Srednicki:1993ix}; $\langle \kappa_{3} \rangle = 0$ and 

\begin{equation}
\label{eq:S2_srednicki}
    \langle \kappa_{3}^2 \rangle = 3 \int_{-1}^1 dx \left( \sum \dfrac{2l+1}{4\pi} C_\ell P_\ell(x) \right)^3 \,,
\end{equation}

\noindent and the solid/dashed green lines are the mean/$\pm 2\sigma$ limits of the Gaussian. Again, we observe that the summary statistic is Gaussianized with increasing $\alpha$, but is highly non-Gaussian for $\alpha < 0$.

In this appendix, we have focused on a hypothetical scenario of an all sky map of a Gaussian random variable and the corresponding one point cumulants. We expect that our conclusions will hold more generally for a summary statistic extracted from a field containing long range correlations. For example, in Figure \ref{fig:ensemble_ngwb_2pfs} we present the PTA two-point function in the 1st (left panel) and 14th (right panel) frequency bins, extracted from 100 pulsars injected with a nanohertz GWB. This corresponds to the ensemble distribution of the sample mean of the two-point function. The histograms were generated from 300 realisations, and different colors denote different correlation function bins. Effectively, these are mock measurements of the binned HD curve from mock PTA data. This shows that the correlation function is strongly non-Gaussian in each angular bin. This is a manifestation of the large scale correlation in the field generating summary statistics that are non-Gaussian.  

We expect our conclusion to hold for any statistic that is a non-linear function of the underlying cosmological fields. This includes the $N$-point correlation functions, Minkowski Functionals, peak statistics, etc. The nature of the probability distribution of these summary statistics is strongly sensitive to the presence of large scale power.

% \bibliographystyle{JHEP}
% \bibliography{refs}

\providecommand{\href}[2]{#2}\begingroup\raggedright\endgroup

\end{document}